\documentclass[acmsmall,screen,authorversion,nonacm]{acmart}

\usepackage{bm}
\usepackage{soul}
\usepackage{prettyref}
\usepackage[most]{tcolorbox}
\usepackage[framemethod=TikZ]{mdframed}
\usepackage{url}

\newcommand{\pref}{\prettyref}
\def\our{\texttt{GraphFLA}}

\definecolor{SoftSkyBlue}{RGB}{220,228,247}
\definecolor{NavyBlue}{RGB}{8,29,92}
\definecolor{greenstar}{RGB}{83, 182, 85}
\definecolor{purplestar}{RGB}{96, 55, 148}
\definecolor{reference}{RGB}{4, 20, 110}
\definecolor{linenum}{RGB}{21, 127, 127}
\definecolor{amaranth}{rgb}{0.9, 0.17, 0.31}
\definecolor{brightmaroon}{rgb}{0.76, 0.13, 0.28}
\definecolor{mybackground}{RGB}{241, 238, 252}
\definecolor{mybackground2}{RGB}{239, 246, 246}
\definecolor{mybackground3}{RGB}{236, 242, 254}
\definecolor{positive}{RGB}{217, 117, 56}
\definecolor{negative}{RGB}{77, 141, 155}
\definecolor{myblue}{RGB}{41, 92, 160}
\definecolor{mygreen}{RGB}{35, 166, 59}
\definecolor{mypurple}{RGB}{96, 54, 147}

\AtBeginDocument{%
  }

\begin{document}

\title[Rethinking Performance Analysis for Configurable Soft. Sys.: A Case Study from a Fitness Landscape Perspective]{Rethinking Performance Analysis for Configurable Software Systems: A Case Study from a Fitness Landscape Perspective}


\author{Mingyu Huang}
\affiliation{%
  \institution{University of Electronic Science and Technology of China}
  \city{Chengdu}
  \country{China}}
\email{m.huang.gla@outlook.com}

\author{Peili Mao}
\affiliation{%
  \institution{University of Electronic Science and Technology of China}
  \city{Chengdu}
  \country{China}}
\email{peili.z.mao@gmail.com}

\author{Ke Li}
\affiliation{%
 \institution{University of Exeter}
 \city{Exeter}
 \country{United Kingdom}}
\email{k.li@exeter.ac.uk}

\renewcommand{\shortauthors}{Trovato et al.}

\begin{abstract}
\footnote{Accepted as a conference paper at \textit{ISSTA 2025}. Please note that this is the peer-reviewed version and not the final camera-ready one, as further modifications are expected.}Modern software systems are often highly configurable to tailor varied requirements from diverse stakeholders. Understanding the mapping between configurations and the desired performance attributes plays a fundamental role in advancing the controllability and tuning of the underlying system, yet has long been a dark hole of knowledge due to its black-box nature. While there have been previous efforts in performance analysis for these systems, they analyze the configurations as isolated data points without considering their inherent spatial relationships. This renders them incapable of interrogating many important aspects of the configuration space like local optima. In this work, we advocate a novel perspective to rethink performance analysis---modeling the configuration space as a structured ``landscape''. To support this proposition, we designed \our, an open-source, graph data mining empowered fitness landscape analysis (FLA) framework. By applying this framework to $86$M benchmarked configurations from $32$ running workloads of $3$ real-world systems, we arrived at $6$ main findings, which together constitute a holistic picture of the landscape topography, with thorough discussions about their implications on both configuration tuning and performance modeling. 
\end{abstract}

\keywords{Configurable software system, Performance analysis, Fitness landscape}

\maketitle

%!TeX root=../main.tex

\section{Introduction}
\label{sec:introduction}

Modern software systems have become increasingly sophisticated and highly configurable, offering numerous tunable options—for example, the Linux kernel has over $15,000$ options. By tuning these \textcolor{NavyBlue}{\textit{configuration options}}, stakeholders are empowered with the flexibility to tailor a system to meet specific functional requirements. Yet on the other hand, these configuration choices can also, intentionally or unintentionally, impact various non-functional \textcolor{NavyBlue}{\textit{performance}} attributes such as execution time and memory consumption of the system~\cite{XuJFZPT15,SiegmundGAK15}. As a result, stakeholders are often overwhelmed by the complexity of the system under tuning and may resort to default settings, which may in turn lead to sub-optimal system behaviors. For example, \cite{HanY16} shows that $59\%$ configuration-related issues can cause severe performance concerns in modern configurable software systems. Therefore, a thorough understanding of the intricate relationship between configurations and system performance is essential for effective system control and has been a long-standing goal in software engineering~\cite{SiegmundGAK15,HaZ19,GuoCASW13,SarkarGSAC15,GuoYSASVCWY18,MuhlbauerSKDAS23,JamshidiSVKPA17,ZhangGBC15,ChengGZ23,DornAS20,Gong024}. However, achieving this understanding in practice is far from trivial due to the black-box nature of the configuration-performance mapping.

\begin{figure}[t!]
    \centering
    \includegraphics[width=\linewidth]{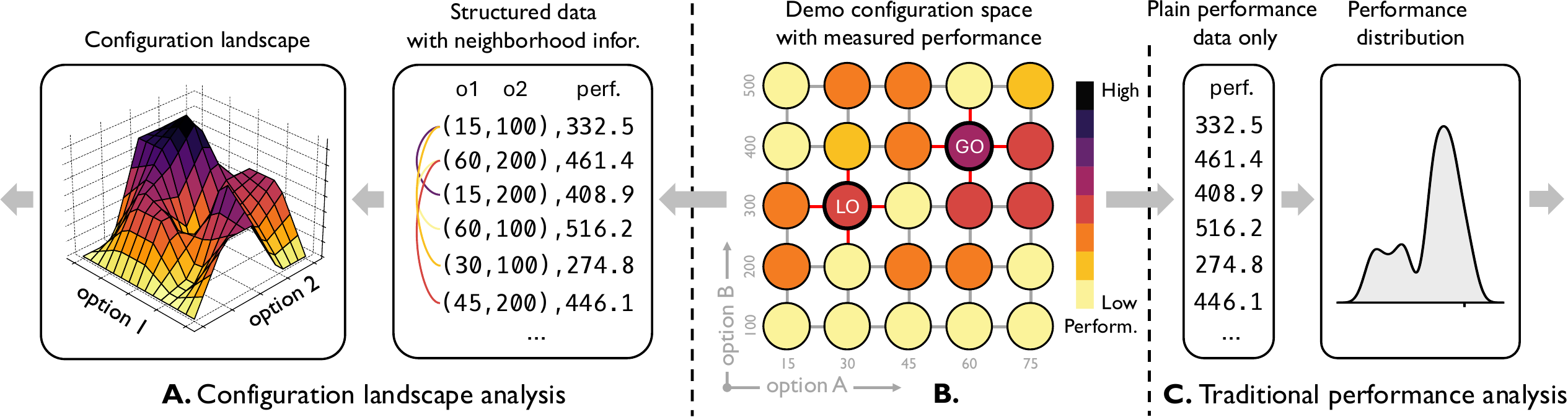}
    \caption{In software configuration space (panel \textbf{B}), configurations are spatially related to each other, and so are their associated performance values. Traditional performance analysis (panel \textbf{C}) only considers the distribution of performance values as isolated data points. Our approach (panel \textbf{A}), instead, additionally incorporates the neighborhood relationships between configurations, which are used to construct a configuration landscape that reveals the spatial distribution of performance values across the configuration space.}
    \label{fig:intro}
\end{figure}

To shed light on these black boxes, empirical \textcolor{NavyBlue}{\textit{performance analyses}} have been used to analyze the distribution of system performance across large configuration spaces and different scenarios~\cite{MuhlbauerSKDAS23,JamshidiSVKPA17}. While performance distributions can provide useful information regarding the system—such as trends, variability, skewness, and kurtosis—they typically consider the performance values of sampled configurations as a set of isolated data points (see panel C of~\pref{fig:intro}). Inherently, however, each performance value is associated with one or more system configurations and different configurations are spatially related to each other within the configuration space (panel B of~\pref{fig:intro}). As a result, the associated performance values also exhibit spatial distributions across the configuration space, in addition to their one-dimensional numerical distributions. 

The lack of such spatial information in performance analysis leaves many important aspects of the software configuration problem unexplored. For example, local optima can present challenges for configuration tuning~\cite{ChenL23a, ZhouHS024}. However, they are not directly observable from performance distributions since this analysis requires examining the spatial adjacency between configurations. Consequently, current analytical methods are unable to interrogate questions like \textit{``are performance local optima prevalent in the configuration space of a given system?''} Similarly, a large variance in overall performance distributions does not necessarily imply notable performance fluctuations in local neighborhoods. Without spatial information, performance distributions alone cannot reveal \textit{``whether the search landscape is smooth or rugged?''} Further, practitioners are often interested in the top-performing configurations. Though prominent performance values can be identified from performance distributions, it is not clear \textit{“how are the corresponding configurations distributed in the configuration space?”} These then lead to our central proposition: 

\begin{tcolorbox}[breakable, title after break=, height fixed for = none, colback = mybackground2, boxrule = 0pt, sharpish corners, top = 4pt, bottom = 4pt, left = 4pt, right = 4pt, toptitle=2pt, bottomtitle=2pt]
    By taking into account the spatial information regarding the configuration space, we can potentially unlock many new insights into the configuration-performance mapping of configurable software systems that were previously inaccessible through traditional performance analyses.
\end{tcolorbox}

Bearing this in mind, we propose a new perspective for performance analysis of configurable software systems by jointly considering the spatial relationships between configurations and their associated performance values. To this end, we leveraged the \textcolor{NavyBlue}{\textit{fitness landscape}} metaphor, a fundamental concept in the evolutionary biology~\cite{VisserK,AguilarPW17,VaishnavBMYFATLCR22,PapkouGEW23} pioneered by Wright nearly a century ago~\cite{Wright32}. In the evolutionary context, ``fitness'' refers to the ability of a phenotype, expressed by a given genotype, to survive and reproduce in a given environment, and the ``landscape'' describes the distribution of all fitness values across the entire genotype space. The resulting fitness landscape contains, in principle, all the information regarding the genotype-fitness mapping (e.g., the number and locations of performance peaks). More generally, this metaphor has been widely adapted for understanding the variable-objective mapping of black-box systems across various domains like metaheuristics~\cite{HuangL23, HuangL23b, Tayarani-NajaranP14, Prugel-BennettT12}, physical chemistry~\cite{Doye02,WangV03} and machine learning (ML)~\cite{PushakH22,MohanBDL23,HuangL23c}.

In the configurable software context, a fitness landscape can be conceived as a (hyper-)surface as formed by system performance values across a high-dimensional configuration space (panel A of~\pref{fig:intro}). Each point in this landscape represents a configuration, with its elevation indicating the fitness and similar configurations are located close to each other. Using this metaphor:
\begin{itemize}
    \item The goal of configuration optimization is then analogous to moving uphill on this landscape toward the global peaks
    \item The goal of performance modeling is equivalent to obtaining an accurate approximation of this surface given a few observations
\end{itemize}
Therefore, knowledge regarding this landscape is a fundamental prerequisite for in-depth understanding of both configuration optimization and performance modeling.

Unfortunately, despite extensive literature on \textcolor{NavyBlue}{\textit{fitness landscape analysis} (FLA)} over decades~\cite{Malan21}, existing methods are not readily accessible or applicable within the software engineering community. A major hurdle is the limited accessibility of these methods. The exploratory nature of FLA often necessitates the use of multiple techniques for a complete understanding of the landscape. Yet, many FLA techniques lack standardized implementations, and those available are developed independently using different programming languages and requiring distinct input data formats. These inconsistencies pose challenges in data preparation, scalability, and the integration of methods into a streamlined pipeline that can be easily employed by software engineers.

To bridge this gap, we present \our, an end-to-end graph-based FLA framework in Python, which enables exploratory analysis of software configuration landscapes with diverse FLA methods and thereby enhances the understanding of the relationship between system configurations and the resulting performance. The \our\ package is purpose-built to construct, analyze, visualize, and compare complex configuration landscapes in a streamlined pipeline and can be applied to software systems of different types, complexities, and domains. At the core of \our\ is an attributed graph representation of the configuration landscape, where each node represents a configuration with performance values as node attributes, and edges connect similar configurations. This graph model naturally incorporates the spatial adjacency between configurations and allows many atomic operations in FLA to be implemented via heavily optimized graph mining algorithms, thus enables it to scale to millions of configurations. Building upon this graph model, \our\ offers a comprehensive suite of FLA techniques that quantify and visualize different aspects of the landscape topography, from which custom analysis pipelines can be built. Its modular architecture further allows users to easily customize the analysis pipeline by plugging in new analysis methods.

To demonstrate how \our\ can deepen the understanding of software configuration problems, we conducted a large-scale case study using three representative real-world configurable systems. With $6$ months of performance benchmarking, we collected over $86$M entries of performance data across $32$ workloads for these systems by combinatorially enumerating configurations formed by $15\sim20$ tunable options. We then applied \our\ to construct high-resolution configuration landscapes from this data and performed extensive analysis on them following a top-down pipeline. Specifically, we began by exploring the overall performance distributions (\pref{sec:f1}), then narrowed down to the most prominent regions (\pref{sec:f2}). Subsequently, we took a more granular examination of particular configurations of interest—local optima and the global optimum—in these landscapes (\pref{sec:f3} and \pref{sec:f4}). Finally, we went beyond the landscape surfaces and explored the individual and interactive performance effects of the studied configurable options (\pref{sec:f5}, \pref{sec:f6}), which are the driving forces shaping the landscape topography. For each subsection, we highlighted the main findings and discussed their implications for both configuration tuning and performance modeling. We ended our analysis in~\pref{sec:alg} by illustrating how these findings can be used to explain algorithmic behaviors. In summary, our main contributions include:

\begin{itemize}
    \item This paper aims to embark on a novel pathway for understanding the intricate configuration-performance mapping of configurable software systems---via a fitness landscape perspective. 
    \item In support of this proposition, we presented \our, an end-to-end graph-based FLA framework that is highly scalable, accessible, and flexible (\pref{sec:graphfla}).
    \item With \our, we conducted the first large-scale case study on software configuration landscapes on $32$ workloads of $3$ real-world systems (\pref{sec:setup}), and systematically mapped and compared diverse aspects of their topographies. This leads to $6$ findings in~\pref{sec:results} with discussions on their implications for both performance optimization and modeling.
    \item We also open-sourced the collected performance data ($6$ months, $3$ systems, $32$ workloads, $86$M$+$ configurations), which can serve as a valuable testbed for many future works.
\end{itemize}

\section{Background and Related Work}
\label{sec:related_works}

\vspace{.3em}
\noindent
\textbf{Configurable software systems.}  A configurable software system is an umbrella term for any kind of software system $\mathcal{S}$ that exhibits configuration options $\{c_1, \dots, c_n\}$ to customize its functionality~\cite{ApelBKS13}, where each configuration option can take either categorical (boolean) or numerical values. Let $\mathcal{C}_i$ be the set of all feasible values for option $c_i$, the whole configuration space $\mathcal{C}$ is then the Cartesian product of the domains of all options, i.e., $\mathcal{C} = \mathcal{C}_1 \times \dots \times \mathcal{C}_n$. While the primary purpose of configuration options is to tune functionality, each configuration choice $\bm{c} = (c_1, \dots, c_n) \in \mathcal{C}$ may also have implications on non-functional performance properties. We denote the performance aspect of interest as $f: \mathcal{C} \to \mathbb{R}$, which is also referred to as the \textit{fitness function} in the FLA context.

\vspace{.3em}
\noindent
\textbf{Software configuration optimization.} For a given software system $\mathcal{S}$, desired performance objective $f$, as well as a configuration space $\mathcal{C}$, the goal of software configuration optimization (tuning) is to find an optimal configuration $\bm{c}^\ast$ that yields the best performance. Formally:
\begin{equation}
    \bm{c}^\ast = {\mathrm{argmin}}_{\bm{c}\in\mathcal{C}}~f(\bm{c}).
\end{equation}
This is a typical $\mathcal{NP}$-hard black-box optimization problem~\cite{Weise09}, and is challenging in several ways: $i)$ by ``black-box'', it means that there is no analytical form of $f$, nor other information like the gradients or the Hessian; $ii)$ the size of $\mathcal{C}$ grows exponentially with the number of options and is thus often astronomical or even infinite; and $iii)$ evaluating $f$ involves actually running the software system, which is costly. Thus, brute-force search is usually infeasible, and efficient tuning of configurable software systems has been rigorously studied for decades under search-based software engineering (SBSE)~\cite{HarmanMZ12}. Common methods for this include random search~\cite{OhBMS17,YeK03}, hill-climbing~\cite{LiZMTZBF14,XiLRXZ04}, genetic algorithms~\cite{BehzadLHBPAKS13,ShahbazianKBM20}, simulated annealing~\cite{DingLQ15}, or Bayesian optimization~\cite{Nair0MSA20,DingPKH20}. 

\vspace{.3em}
\noindent
\textbf{Software performance modeling.} Apart from configuration optimization, another related line of research attempts to develop a surrogate function $\hat{f}$ that maps from $\mathcal{C}$ to $f$ by learning from sampled configurations. This can be achieved by different ML models, e.g., from the probabilistic programming~\cite{DornAS20} and linear regression~\cite{SiegmundGAK15} to Fourier learning~\cite{ZhangGBC15,Ha019} and tree-based models~\cite{GuoCASW13,GuoYSASVCWY18,NairMSA17,SarkarGSAC15}. More recently, deep neural networks (DNNs) have also gained increasing popularity in this domain with prominent performances~\cite{GongC24,HaZ19}. Without loss of generality, performance models come with a set of parameters $\theta$ to be fitted to the training data, and the goal of the training process is to minimize the loss function between the predicted performance and the actual performance of configurations. The mean squared error (MSE) below is one possible choice for the loss function:
\begin{equation}
    \hat{\theta} = \underset{\theta}{\mathrm{argmin}} \frac{1}{m} \sum_{i=1}^{m} \left(f(\bm{c}_i) - \hat{f}(\bm{c}_i; \theta)\right)^2.
\end{equation}

\vspace{.3em}
\noindent
\textbf{Software configuration landscape.} While historically FLA mainly concerns biological systems and test problems in metaheuristics, here we adapt this notion to define software configuration landscapes. The configuration landscape for software systems is defined as a triplet $(\mathcal{C}, \mathcal{N}, f)$, where $\mathcal{N}$ is a neighborhood structure that specifies which configurations are neighbors to each other based on a distance function (e.g., Hamming distance). Intuitively, a landscape can be seen as a structured set of $\langle \mathbf{c}_i, f(\mathbf{c}_i) \rangle$ pairs, where each $\mathbf{c}_i$ is related to other $\mathbf{c} \in \mathcal{C}$ as indicated by $\mathcal{N}$. Over the past decades, numerous FLA techniques have been developed in both evolutionary biology and metaheuristics to characterize the topography of this surface~\cite{VisserK,MalanE13,Malan21}. Though a comprehensive review is beyond our scope, we introduce some basic concepts that are relevant to our work: 
\begin{itemize}
    \item \textbf{Diameter:} The diameter of a landscape is the longest distance between any two configurations in $\mathcal{C}$. The radius of a landscape is defined accordingly as half of the diameter.
    \item \textbf{Local optima, landscape modality and ruggedness:} A configuration $\bm{c}^\ell$ is said to be a local optimum iff $f(\bm{c}^\ell)$ is better than ${f(\bm{c})}$, $\forall\bm{c}\in{\mathcal{N}(\bm{c}^\ell)}$. A landscape with many local optima is considered to be rugged, and multimodal. 
    \item \textbf{Epistasis:} In biology, epistasis refers to non-additive interactions between genes, where the combined effect differs from the sum of individual effects. In our software configuration context, this phenomenon is known as feature interactions\footnote{To enhance readability for general audience, we use feature interaction instead of epistasis in the rest of this paper.}—where the impact of one option on system performance depends on the values of others. Feature interaction is a key factor that shapes the ruggedness of a landscape—In the simplest case of no interactions between options (i.e., $f$ consists of only additive terms), the landscape would be fairly smooth with a single optimum, which can be reached by optimizing each option independently. In contrast, when interactions are present, the landscape becomes rugged with many local optima.
\end{itemize}

\vspace{.3em}
\noindent
\textbf{Software performance analysis.} Also related to our work are those performing traditional performance analysis on configurable software systems~\cite{JamshidiSVKPA17,MuhlbauerSKDAS23,JamshidiVKS18,MartinAPLJK22,ValovPGFC17}. While these literature share some techniques with ours, such as performance distribution and performance-influence analysis~\cite{Ha019}, our FLA perspective places these analyses into a much broader context (see our FLA pipeline in~\pref{fig:diagram}). Also, as demonstrated in~\pref{sec:f5}, our FLA approach enables more accurate, granular analysis of the performance effects of individual options, which are difficult to achieve with traditional performance-influence models. Beyond these overlaps, \our\ allows us to explore additional critical aspects of the software configuration space, such as prominent regions (\pref{sec:f1}), local optima, and global optimum (\pref{sec:f3} and \pref{sec:f4}), which were previously inaccessible due to the lack of spatial information. Apart from these methodological differences, the absence of spatial information in traditional performance analysis also constrains the interpretation primarily to performance modeling. In contrast, our investigation of landscape topographies has a direct impact on both performance modeling and configuration tuning. Finally, while there is also extensive work on the white-box analysis of configurable software systems (e.g., \cite{VelezJSSAK20,VelezJSAK21,WeberAS21,HanYP21}), our approach is complementary to them, as we focus on the black-box perspective.

% \begin{itemize}
%     \item  
%     \item \textbf{Local optimum.} A configuration $\bm{c}^\ell$ is said to be a \textit{local optimum} iff $f(\bm{c}^\ell)$ is better than ${f(\bm{c})}$, $\forall\bm{c}\in{\mathcal{N}(\bm{c}^\ell)}$. Here $\mathcal{N}(\bm{c}^\ell)$ is the neighborhood of $\bm{c}^\ell$.
%     % \item \textbf{Basin of attraction.} The \textit{basin of attraction} of a local optimum $\bm{c}^\ell$, denoted as  $\mathcal{B}(\bm{c}^\ell)$, is the set of all configurations from which local search converges to it. We define the \textit{size} of a basin to be its cardinality $|\mathcal{B}|$, and its \textit{radius} to be the expected number of local search steps required to arrive at the local optimum. 
%     %Depending on the local search strategy used to define it, e.g., \textit{best-} or \textit{first-}improvement~\cite{WhitleyHH13}, the basin of a local optimum can vary.
%     % \item \textbf{Epistasis.} It is initially used in evolutionary biology to describe the suppression or enhancement of gene expression at one locus by another gene, and is analogous to the term \textit{feature interaction} here in software engineering.
    
% \end{itemize}

%!TeX root=main.tex

\section{The FLA Framework and Case Study Setup}
\label{sec:methods}

\subsection{GraphFLA}
\label{sec:graphfla}

\begin{figure}[t!]
    \centering
    \includegraphics[width=\linewidth]{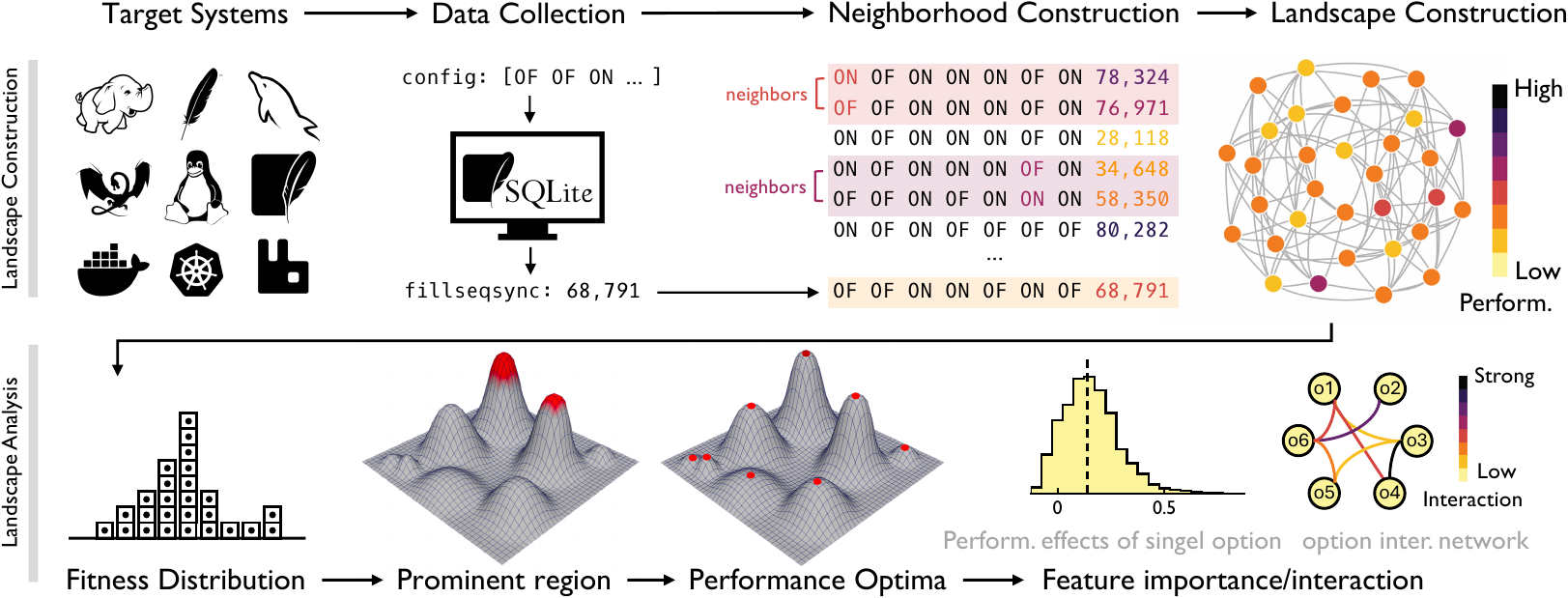}
    \caption{Schematic overview of of \texttt{GraphFLA}.}
    \label{fig:diagram}
\end{figure}

As illustrated in~\pref{fig:diagram}, the foundation of \our\ is a versatile graph object for representing configuration landscape, which is constructed from sampled configuration data and specified neighborhood structure, and then analyzed using a suite of FLA methods.

\subsubsection{Data collection.} Given an arbitrary software system $\mathcal{S}$, configuration space $\mathcal{C}$ and target performance objective $f$, the first step is to obtain a set of representative data points $\{\langle\mathbf{c},f(\mathbf{c})\rangle\}$ from $\mathcal{C}$. To allow for standardized analysis of performance data and interoperability with existing benchmarks, \our\ accepts pre-collected data in consistent, reusable formats, e.g., CSV, JSON, or SQL, etc. In addition, \texttt{GraphFLA} also provides support to various data sampling methods that can be used in conjunction with performance benchmarking tools to collect data points on the fly:

\begin{itemize}
    \item \textbf{Exhaustive grid search:} This involves enumerating all possible configurations in $\mathcal{C}$ and evaluating their performance. While straightforward and expensive, this approach offers the highest resolution and coverage of the landscape. It also enables better interpretability of the resulting landscape---neighboring configurations only differ by exactly one option and thereby allows us to accurately analyze the individual performance effect of each option. Bearing these considerations in mind, we employed such a grid search for this empirical study, and we will discuss the details later in~\pref{sec:setup}.
    \item \textbf{Sampling methods:} In larger configuration spaces where an exhaustive search is unfeasible, sampling methods can be employed in \our\ to collect a representative set of data points, which enables a cost-effective sketch of the landscape topography. Moreover, performance models can be used in conjunction with sampled data points to improve coverage and resolution without additional performance benchmarking efforts. They have been successfully applied in other domains for landscape interpolation~\cite{SiemsZZLKH20,VaishnavBMYFATLCR22,RodriguezTAB24}.
\end{itemize}

\subsubsection{Distance measures and neighborhood structure.} After collecting the configuration-performance data, recap from the definition in \pref{sec:related_works} that we still need a third ingredient for constructing the landscape: a neighborhood structure $\mathcal{N}$. Central to this is a proper definition of distance $d(\bm{c}_i,\bm{c}_j)$ between configurations. As one can imagine, the choice of this distance measure is closely related to the nature of the configuration space. For example, distances between categorical options are better represented by the Hamming distance, whereas for ordinal values, the Euclidean distance or Manhattan distance could be more appropriate. For more complex scenarios, e.g., when hierarchical or relational structures are involved, more sophisticated distance measures like the tree-edit distance~\cite{PawlikA15} can be used. Fortunately, in this study, we do not have such complex search spaces (\pref{sec:setup}). Therefore, we went with the Hamming and Manhattan distances, for categorical and integer options respectively. Given this, the total distance $d(\bm{c}_i,\bm{c}_j)$ between two configurations can be calculated as the sum of the distances for each corresponding pair of values. Following classic FLA literature, the neighborhood $\mathcal{N}(\bm{c})$ of a configuration $\bm{c}$ is then defined as the set of all configurations differing from $\bm{c}$ in exactly one configuration option, i.e., $\mathcal{N}(\bm{c})=\{\bm{c}^\prime\mid d(\bm{c^\prime},\bm{c})=1\}$. 

\subsubsection{Configuration landscape as a graph.} Using the defined neighborhoods and the collected data points, \texttt{GraphFLA} then models the spatial relationship among the configurations as a directed, attributed graph. In this graph model, each node represents a configuration $\bm{c}$, with the performance metric $f(\bm{c})$ assigned as the node attribute. Neighboring configurations, denoted as $\bm{c}^\prime\in\mathcal{N}(\bm{c})$, are connected to $\bm{c}$ via a directed edge. The direction of each edge is determined by the relative values of $f(\bm{c})$ and $f(\bm{c}^\prime)$, always pointing towards the fitter configuration. This graph model then enables the identification of local optima in the landscapes, as well as the implementation of FLA methods, using straightforward graph mining techniques. For example, locating local optima in the landscape can be solved by finding sink nodes in the resulting graph, while atomic FLA operations like random walk, can be readily grounded on its graph counterpart. 

\subsubsection{Configuration landscape analysis.} Powered by graph representation that scales to landscapes with millions of configurations, \our\ provides diverse composable analysis functions organized in modules from which custom analysis pipelines can be built. Each function directly interacts with the landscape object and adds all intermediate results for simple access and reuse of information to it. To facilitate setting up these pipelines, \our\ guides analysis through a general workflow (see~\pref{fig:diagram}). In this pipeline, \our\ explores the configuration landscape using a top-down approach, and we followed exactly the same procedure during our analysis in~\pref{sec:results}. Specifically, It begins by exploring the performance distribution of the configurations, revealing its trend and variability. Subsequently, \our\ identifies prominent regions in the landscapes and tracks their distribution across the landscape, which are the ultimate goals of configuration tuning. Next, it zooms into particular configurations of interest, particularly local optima and global optimum, and examines their quantities and locations, shedding light on the ruggedness of the landscape topography. Beyond these, \our\ can also reveal hidden mechanisms shaping such topography, i.e., the performance effects of individual options and their interactions. 

As analysis goals can differ between users and systems, this pipeline is customizable. At any step of an analysis pipeline, other community software packages or new FLA functions can be easily plugged in. \our\ also provides statistical methods for group comparisons between landscapes and interactive visualization tools to facilitate the interpretation of landscape topography. Moreover, by cooperating with ML libraries like scikit-learn/PyTorch or optimization platforms like Optuna~\cite{AkibaSYOK19}, \our\ can serve as a testbed for analyzing model performance or algorithmic behavior.

\subsection{Setup for Our Empirical Case Study}
\label{sec:setup}

\begin{table}[t!]
    \caption{Meta-information of our experiments.}
    \vspace{-0.3cm}
    \centering
    \small
    \begin{tabular}{||lcccccc|c||}
        \hline
        System & Language & Domain & Previously Considered by & $|\mathcal{W}|$ & $|\mathbf{c}|$ & $|\mathcal{C}|$ & Total Eval. \\
        \hline
        \hline
        \textsc{LLVM}   & C++     & Compiler    & \cite{GuoCASW13,SarkarGSAC15,SiegmundKKABRS12,OhBMS17,KolesnikovSKGA19,Nair0MSA20,KalteneckerGSGA19,GuoYSASVCWY18,SiegmundGAK15}   & $12$ & $20$ & $1.05$M & $12.58$M \\
        \textsc{Apache} & C       & Web Server  & \cite{GuoCASW13,GuoYSASVCWY18,NairMSA17,SarkarGSAC15,SiegmundKKABRS12,JamshidiSVKPA17,ValovPGFC17,NairMSA18,TempleAJB17}  & $10$ & $18$ & $1.77$M & $17.69$M \\
        \textsc{SQLite} & C       & Database    & \cite{GuoCASW13,GuoYSASVCWY18,OhBMS17,KolesnikovSKGA19,NairMSA17,SarkarGSAC15,SiegmundKKABRS12,YilmazDCP14,NairMSA18}   & $10$ & $16$ & $5.67$M & $56.65$M \\
        \hline
        \hline
        \textsc{Total}: &         &          &          & $\bm{32}$ &  &  & $\bm{86.92\textbf{M}}$ \\
        \hline
    \end{tabular}
    \label{tab:setup}\\
    \vspace{0.1cm}
    \raggedright\fontsize{8pt}{8pt}\selectfont $|\mathcal{W}|$ is the number of workloads; $|\mathbf{c}|$ is the number of selected options; $|\mathcal{C}|$ is the total size of the configuration space.
\end{table}

\subsubsection{Subject systems, options, and workloads.} As a case study, with a focus on external validity, we selected three highly configurable systems that have been widely used in literature (see~\pref{tab:setup}). These systems are from different domains in which configuration tuning is important, and written in different programming languages. They are also of different sizes (lines of code) with different numbers of configuration options. For each system, we generated a diverse set of workloads, and thereby constituted a total of $32$ different scenarios. Note that ideally, each set of workloads is diverse enough to be representative of most possible use cases. We selected the workload sets in this spirit, but cannot always guarantee a measurable degree of diversity and representativeness prior to the actual measurements. We will discuss this as a threat to validity.

\begin{itemize}
    \item \textbf{\textsc{LLVM}}: The \textsc{LLVM} Project is a collection of modular compiler and toolchain technologies. It has $578$ configuration options and we chose $20$ of them for our empirical study\footnote{We employed a two-stage feature selection process to select a subset of configurable options for each system: Initially, we conducted a pre-selection based on the official documentation, including only options explicitly stated to affect performance. We then performed an ablation analysis~\cite{BiedenkappLEHFH17} to filter out options without statistically significant impact on performances.}. 
    We generated different workloads by compiling $12$ different program\footnote{The complete information regarding the selected workloads and options are available in \href{https://zenodo.org/records/14021213}{Appendix}.} codes from the \texttt{PolyBench} benchmark suite, covering diverse application domains such as algebra (\texttt{2mm}), image processing (\texttt{deriche}), data mining (\texttt{correlation}). They also span a wide range of computational patterns, memory access behaviors, and algorithmic complexities. The run time for the compiled program is used as the fitness function for each configuration of \texttt{LLVM}.
    \item \textbf{\textsc{Apache}}: The \textsc{Apache} HTTP server project aims to offer a robust and scalable HTTP service.  It consists of multiple modules, the core of which has $89$ configuration options and $21$ configuration options for MPM module. Here we choose $15$ options directly related to the quality of a configuration in our experiments. Using the Apache HTTP server benchmarking tool, we created $9$ workloads by varying two parameters: \texttt{requests} (total number of requests) and \texttt{concurrency} (number of concurrent requests). These parameters were varied systematically from low to high values (\texttt{requests} from $50$ to $1000$ and \texttt{concurrency} from $50$ to $500$) to simulate a wide range of real-world conditions, including light, moderate, and heavy server loads. We used the requests handled per second as the fitness function, a widely recognized performance measure in web server studies.
    \item \textbf{\textsc{SQLite}}: This is an embedded database project.  It has $50$ compile-time and $29$ run-time configuration options and we chose $18$ of them in this study. The Google-provided SQLite Benchmark was used to create $10$ workloads. Each workload was generated by varying two parameters: \texttt{num} (number of entries) and \texttt{value\_size} (size of each entry), which affect the size and complexity of data management tasks. These workloads were constructed to represent database environments with differing sizes and data complexity, from small, simple databases to larger, more demanding ones. The writing speed in sequential key order in async mode is used as the fitness function.
\end{itemize}

\subsubsection{Data collection.} For each system and workload, we exhaustively explored all combinations of categorical options and used a grid search for integer options (See \href{https://zenodo.org/records/14021213}{Appendix} for the adopted search space). This resulted in $1.7$M configurations for each workload of \textsc{LLVM}, $2.9$M for \textsc{Apache}, and $5.0$M for \textsc{SQLite}. Across $32$ workloads from the $3$ systems, the total number of configurations considered was over $86$M, which is, to our knowledge, the largest scale in the literature. For each configuration, we evaluated its performance with $10$ independent runs to mitigate the impact of measurement bias~\cite{MytkowiczDHS09}, and we used the mean for this study. During this process, configuration options that were not considered were set to their recommended values. To minimize measurement noise, we used a controlled environment, where no additional user processes were running in the background, and no other than necessary packages were installed. We ran each subject system exclusively on separate nodes of a cluster (see below).

\subsubsection{Hardware platform.} All performance benchmarking tasks in this paper were run on a cluster with $20$ nodes, each equipped with an Intel$^{\text{\textregistered}}$ Core$^{\text{TM}}$ i$7$--$8700$ CPU@$3.10$GHz, 16GB memory and run a CentOS 7.9.2009 distribution. Evaluating all $86$M configurations from the $3$ systems with $10$ repetitions took about $6$ months to complete, which results in a total of more than $86,400$ CPU hours. For the subsequent landscape construction and analyses, all the experiments were carried out using a single node with Intel$^{\text{\textregistered}}$ Xeon$^{\text{\textregistered}}$ Platinum $8260$ CPU@$2.40$GHz and $256$GB memory.

\section{Configuration Landscape Analysis}
\label{sec:results}

We now apply the FLA pipeline introduced in~\pref{sec:graphfla} to analyze the constructed landscapes, which break down into six subsections. At the end of each subsection, we provide a summary of the findings and discuss their implications on both performance optimization and modeling. To further illustrate the impact of landscape topography on algorithmic behaviors, we conducted additional simulation experiments on the constructed landscapes in~\pref{sec:alg}.

\begin{figure}[t!]
    \centering
    \includegraphics[width=\linewidth]{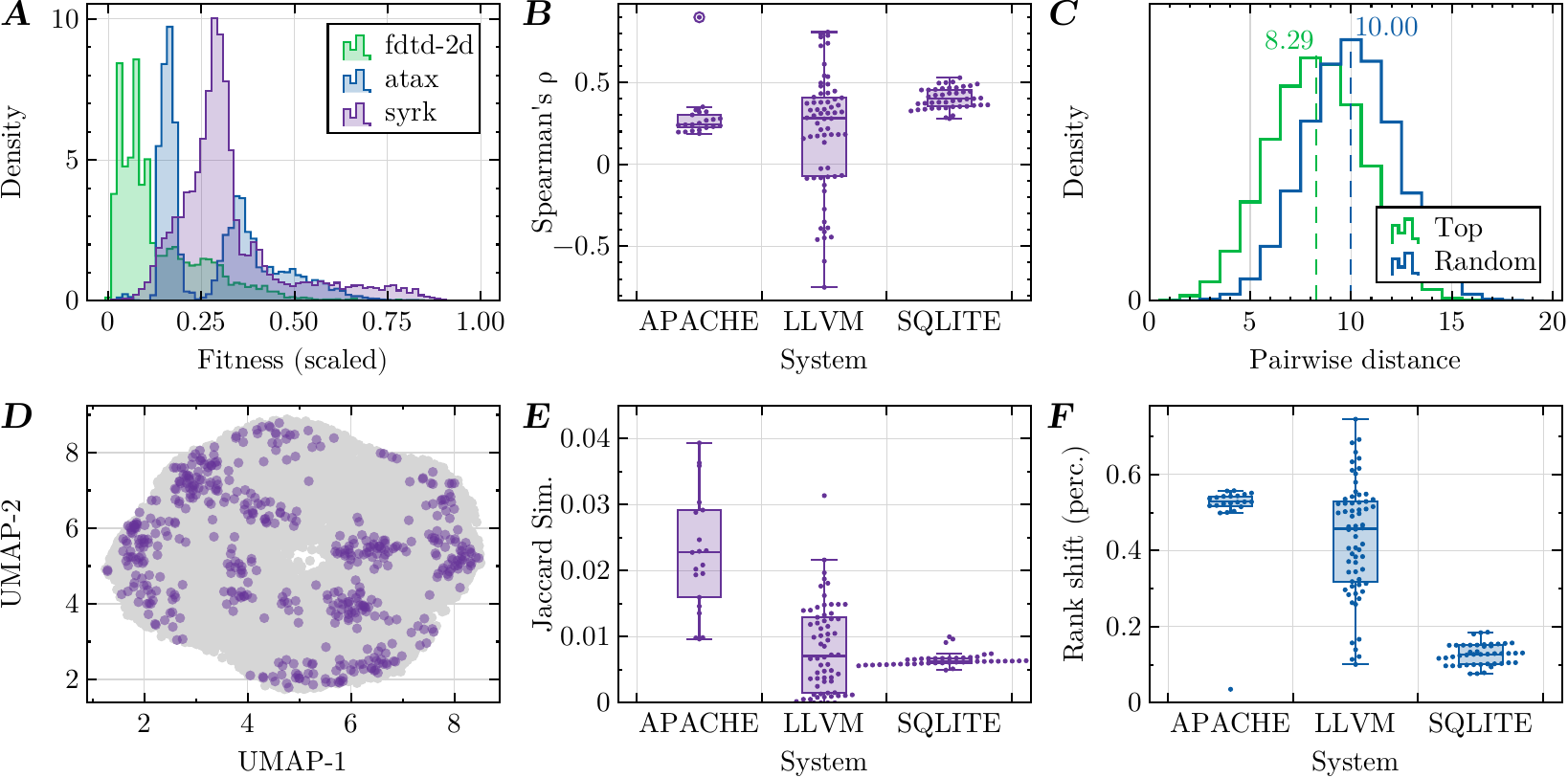}
    \caption{\textbf{General fitness distributions and prominent regions}. \textbf{(A)} Normalized fitness distributions of three workloads of LLVM. \textbf{(B)} Spearman's $\rho$ of fitness between all pairs of workloads in each system. \textbf{(C)} Distributions of pairwise distance between \textcolor{mygreen}{top-$1\%$ configurations} for a \textsc{LLVM} landscape and \textcolor{myblue}{randomly sampled configurations}. \textbf{(D)} 2D projection of the distribution of \textcolor{mypurple}{top-$1\%$ configurations} in a \textsc{LLVM} landscape as compared to total configurations. \textbf{(E)} Overlaps in top-$1\%$ configurations between all pairs of workloads in each system. \textbf{(F)} Average rank shifts (percentile) of fitness for top-$1\%$ configurations between workloads.}
    \label{fig:rq1}
\end{figure}

\subsection{Fitness distribution} 
\label{sec:f1}

Before we delve into the detailed landscape topography, we first took a look at the general fitness distributions. In panel A of~\pref{fig:rq1}, we plotted the normalized fitness distributions of three workloads of \texttt{LLVM}. From the plot, we can see that the distribution for the \texttt{syrk} and \texttt{fdtd-2d} workloads are nearly unimodal, whereas that for \texttt{atax} shows a distinct bimodal pattern. In addition, all three distributions are positively skewed, with a long tail at the right side. Such skewness is also prevalent for other workloads and systems (mean Fisher's skewness $=0.8709$).

To quantify fitness shifts across workloads, a concerned feature for transfer learning~\cite{JamshidiVKSK17,ValovPGFC17,MartinAPLJK22}, we assessed Spearman's correlation between the fitness distributions of each pair of workloads. The results in panel B of~\pref{fig:rq1} indicate that the fitness distributions between most workloads among the three systems are weakly to moderately correlated ($|\rho| < 0.7$). While all correlations for \texttt{SQLite} and \texttt{Apache} are positive, we observed a set of negative correlations for \texttt{LLVM}. These usually occur between \texttt{atax}/\texttt{deriche} and other workloads, which could be classified as severe environmental changes introduced in~\cite{JamshidiSVKPA17}. 

\begin{tcolorbox}[breakable, title after break=, height fixed for = none, colback = mybackground, boxrule = 0pt, sharpish corners, top = 4pt, bottom = 4pt, left = 4pt, right = 4pt, toptitle=2pt, bottomtitle=2pt]
    \textbf{F1:} The fitness distribution of our studied configuration landscapes is highly skewed. Moderate fitness correlations are observed between most workloads, yet their distributions may exhibit distinct patterns.\\
    \textbf{Implications:} Ideally, transfer-based performance modeling or optimization desires a high resemblance between the fitness distributions of the source and target workloads. This constitutes (part of) the key information that many prior works rely on~\cite{JamshidiSVKPA17,JamshidiVKS18}. However, the distinct patterns and moderate correlations observed here suggest that further efforts are needed in the selection of suitable source workloads for transfer. To this end, the integration of domain-specific knowledge like workload characteristics~\cite{CalzarossaMT16}, or learning mechanisms~\cite{WeiZHY18,RuderP17} can be further explored. In addition, \cite{HuangL23} shows that landscape characteristics can also be exploited to reflect the similarities between different problem instances. 
\end{tcolorbox}

\subsection{Prominent regions}
\label{sec:f2}

We next zoomed into the most prominent regions of the landscape, which are of particular interest in practice. In doing so, here we considered the configurations with top-$1\%$ fitness in each landscape. The very first question is whether these configurations would form an interconnected fitness ``plateau'', or they are scattered across the landscape as isolated points. To explore this, we analyzed the distribution of pairwise distances between top-$1\%$ configurations and compared it with that of $10k$ randomly sampled configurations from each landscape. While the results suggest that the top-$1\%$ configurations are far from randomly distributed (two-sided Kolmogorov-Smirnov test $D>23.8$, $p<10^{-23}$, see panel C of~\pref{fig:rq1} for an example), we found that most of these configurations are still quite distant from each other ($d=8.29 \pm 2.13$, mean $\pm$ std). As a more intuitive illustration, we further applied HOPE graph embedding~\cite{OuCPZ016} along with UMAP dimensionality reduction~\cite{McInnesHSG18} method to project the landscape into a 2D space in panel D of~\pref{fig:rq1}. From the plot, we can see that the prominent configurations form several local clusters that are widely scattered across the landscape, which is against the hypothesis of a single interconnected plateau. In fact, a sub-landscape (i.e., a subgraph) formed by only the top-$1\%$ configurations would typically contain thousands of ($3,439 \pm 871$ across $32$ scenarios) independent components.

Another important question is whether these prominent configurations are shared across different workloads of the same system, which is another desired property of transfer learning methods. To this end, we first calculated the Jaccard similarity between the top-$1\%$ configurations of each pair of workloads in a system. We surprisingly found that there is little overlap between these configurations of different workloads, with the Jaccard similarity typically $<0.03$ (panel E of \pref{fig:rq1}). We next determined the average shift in fitness percentile for these configurations between different workloads, for which we found that for \texttt{SQLlite}, directly transferring a top-$1\%$ configuration from one workload to another would result in only $12.7\% \pm 2.7\%$ fitness shift, whereas for \texttt{LLVM} and \texttt{Apache}, the shift could be as high as $43.1\% \pm 14.9\%$ and $50.7\% \pm 10.7\%$, respectively (panel F of \pref{fig:rq1}). For these two systems, sometimes the fitness of a top-$1\%$ configuration in one workload could drop to the bottom $10\%$ in another.

\begin{tcolorbox}[breakable, title after break=, height fixed for = none, colback = mybackground, boxrule = 0pt, sharpish corners, top = 4pt, bottom = 4pt, left = 4pt, right = 4pt, toptitle=2pt, bottomtitle=2pt]
    \textbf{F2:} The top-$1\%$ configurations are scattered across each studied landscape, forming sparse local clusters. There is minimal overlap between these configurations for different workloads of the same system, and significant performance shifts can occur when transferring them directly.\\
    \textbf{Implications:} The little overlap and significant performance shifts in the top-$1\%$ configurations between workloads suggest that directly transferring prominent configurations across workloads may not lead to satisfactory results. Despite this, their widespread can make the initialization of the optimizer less critical, as there is no single ``good'' starting point. Additionally, the existence of clustering patterns of top configurations also implies that it is possible to ``learn'' such information and then leverage this to guide optimization. To this end, \cite{LustosaM24} has made successful attempts, and we believe this is a promising direction for future research. 

\end{tcolorbox}

\begin{figure}[t!]
    \centering
    \includegraphics[width=\linewidth]{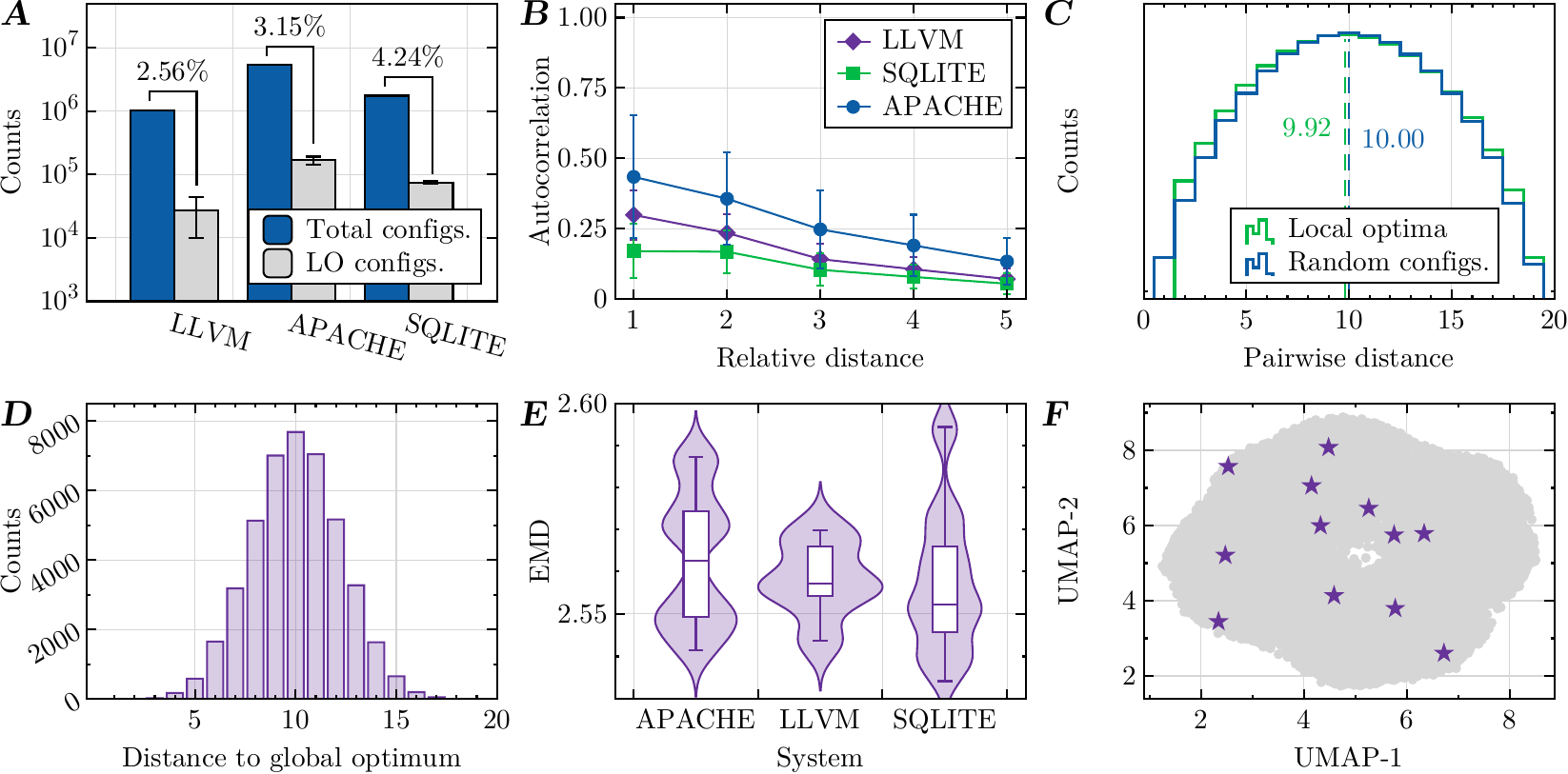}
    \caption{\textbf{Local optima, global optimum, and landscape ruggedness.} \textbf{(A)} Number of local optima and total configurations in landscapes, error bars indicate s.t.d. across workloads. \textbf{(B)} Autocorrelations of landscapes of each system under different lag distances, aggregated across workloads. \textbf{(C)} Exemplar distributions of pairwise distance between all pairs of \textcolor{mygreen}{local optima} in a landscape and the same number of \textcolor{myblue}{randomly sampled configs}. \textbf{(D)} Exemplar distribution of the distance between each local optimum and the global optimum for a landscape. \textbf{(E)} EMD for local optima between all pairs of workloads in each system. \textbf{(F)} 2D projection of the distribution of \textcolor{mypurple}{global optimum} in each of the $12$ LLVM landscapes as compared to total configurations (grey).}
    \label{fig:rq1_2}
\end{figure}

\subsection{Local optima and landscape ruggedness} 
\label{sec:f3}

We now further zoom into specific configurations in the landscape, i.e., local optima, which serve as a coarse-grained indicator of landscape ruggedness. We first quantified the number of local optima in each studied landscape. We found that all of them exhibit a significant number of local optima, ranging from $10^4$ to $10^5$, which take on average $2.56\%$ (\texttt{LLVM}) to $4.24\%$ (\texttt{SQLite}) of the total configurations (panel A of~\pref{fig:rq1_2}). This is comparable to a maximally rugged ($k=19$) Kauffman's $NK$ landscape~\cite{Kauffman93} of dimension $n=20$ (proportion of local optima being $\sim4.43\%$). 

Such observed ruggedness is further supported by the autocorrelation~\cite{Weinberger90} measured on each landscape (panel B of~\pref{fig:rq1_2}), which is another widely used method for assessing landscape ruggedness. It is calculated as the autocorrelation of fitness values across configurations visited during random walks through the landscape, and reflects the degree of fitness fluctuations in the landscape. The corresponding results show a high concordance with the large number of local optima, where even for the closest configurations (i.e., $d=1$), their fitness values are only weakly correlated. This correlation diminishes further as the distance increases.

\begin{tcolorbox}[breakable, title after break=, height fixed for = none, colback = mybackground, boxrule = 0pt, sharpish corners, top = 4pt, bottom = 4pt, left = 4pt, right = 4pt, toptitle=2pt, bottomtitle=2pt]
    \textbf{F3:} Studied landscapes are highly rugged, with abundant local optima and fitness fluctuations.\\
    \textbf{Implications:} The abundant local optima in the configuration landscapes can easily trap optimizers. Some strategies have shown promise in counteracting this, e.g., multi-objectivization~\cite{ChenL21}, random restarting~\cite{LiZMTZBF14}, and landscape smoothing~\cite{KirjnerYSBJBF24}, but more efforts are required.
    For performance modeling, the impact of landscape ruggedness on models' performances has been rarely discussed. Yet, as we will show later in~\pref{fig:predictability}, there is a remarkable correlation between the two. How to cope with landscape ruggedness in the modeling paradigm is still an important open question. Potential ways forward can include using ensemble models~\cite{FreschlinFHR24}, data augmentations (e.g., by resampling for extreme values~\cite{ChawlaBHK02}), or employing regularization.
\end{tcolorbox}

\subsection{Distribution of local optima} 
\label{sec:f4}

In landscapes with numerous local optima, it is then crucial to know their spatial distributions, e.g., whether they are locally clustered or widely dispersed. To investigate this, we analyzed the distribution of pairwise distances between local optima and compared it with that of $10k$ randomly sampled configurations from each landscape. Surprisingly, across all scenarios, the two distributions were remarkably similar (two-sided Kolmogorov-Smirnov test $p > 0.954$; see example in panel C of~\pref{fig:rq1_2}). This suggests that local optima are almost uniformly distributed across the landscapes. While this analysis considers distances between all pairs of local optima, we also examined particular regions of interest. For example, panel D of~\pref{fig:rq1_2} depicts the distribution of distances between each local optima and the global optimum in a \texttt{LLVM} landscape, which reveals that most local optima are located far from the global optimum.

Another aspect to explore is whether the local/global optima in landscapes are shared across different workloads of the same system. In other words, if a configuration is known to be a local/global optimum in one workload, is it likely to be a local/global optimum in another workload? To explore this, we first assessed the Jaccard similarity between the set of local optima in each pair of workloads. This yielded similar results as in panel E of~\pref{fig:rq1}, i.e., there is typically only $1\%\sim 3\%$ overlap between local optima of different workloads. Yet, this is a rather strict criterion as it seeks exact matches between configurations. Alternatively, we hypothesized that the local optima in different workloads distribute similarly in the landscape, but undergo slight ``shifts''. To quantify such distributional similarity of local optima, we adopted the earth mover's distance (EMD)~\cite{RubnerTG00} from the classic optimal transport framework. It quantifies the minimal ``work'' (i.e., distance) required to relocate a set of local optima in a landscape into a different distribution (as in a different landscape). The results shown in~\pref{fig:rq1_2}E indicate that despite the low Jaccard similarity, the distribution of local optima in different workloads exhibits close proximity.

As for global optima, however, we found that they tend to be located in very different regions of the landscape across workloads. For example, \pref{fig:rq1_2}F depicts the 2D projection of the distribution of global optima in each of the $12$ \texttt{LLVM} landscapes, where their pairwise distance is $9.44 \pm 2.48$, comparable to the landscape radius $r_{\mathrm{LLVM}=10}$. We observed similar patterns for \textsc{Apache} and \textsc{SQLite} landscapes, where the pairwise distance between global optima is $12.54 \pm 3.67$ ($d_{\mathrm{Apache}=14}$) and $9.64 \pm 2.14$ ($d_{\mathrm{SQLite}=11}$), respectively.

\begin{tcolorbox}[breakable, title after break=, height fixed for = none, colback = mybackground, boxrule = 0pt, sharpish corners, top = 4pt, bottom = 4pt, left = 4pt, right = 4pt, toptitle=2pt, bottomtitle=2pt]
    \textbf{F4:} The local optima in our investigated configuration landscapes are randomly scattered across the landscapes, with most located far from the global optimum. The local optima in different workloads exhibit close proximity in their distribution, but the global optima are usually located in very different regions of the landscape.\\
    \textbf{Implications:} The random distribution of local optima reinforces the previous implications regarding transfer learning discussed in \textbf{F2}. Yet, the close proximity of local optima between workloads suggests that an escaping mechanism that works for one environment will likely work for the others too; given the distant global optima in different workloads, directly reusing such configurations~\cite{KinneerGG21} or via warm-starting the optimization may not be beneficial.  
\end{tcolorbox}

\begin{figure}[t!]
    \centering
    \includegraphics[width=\linewidth]{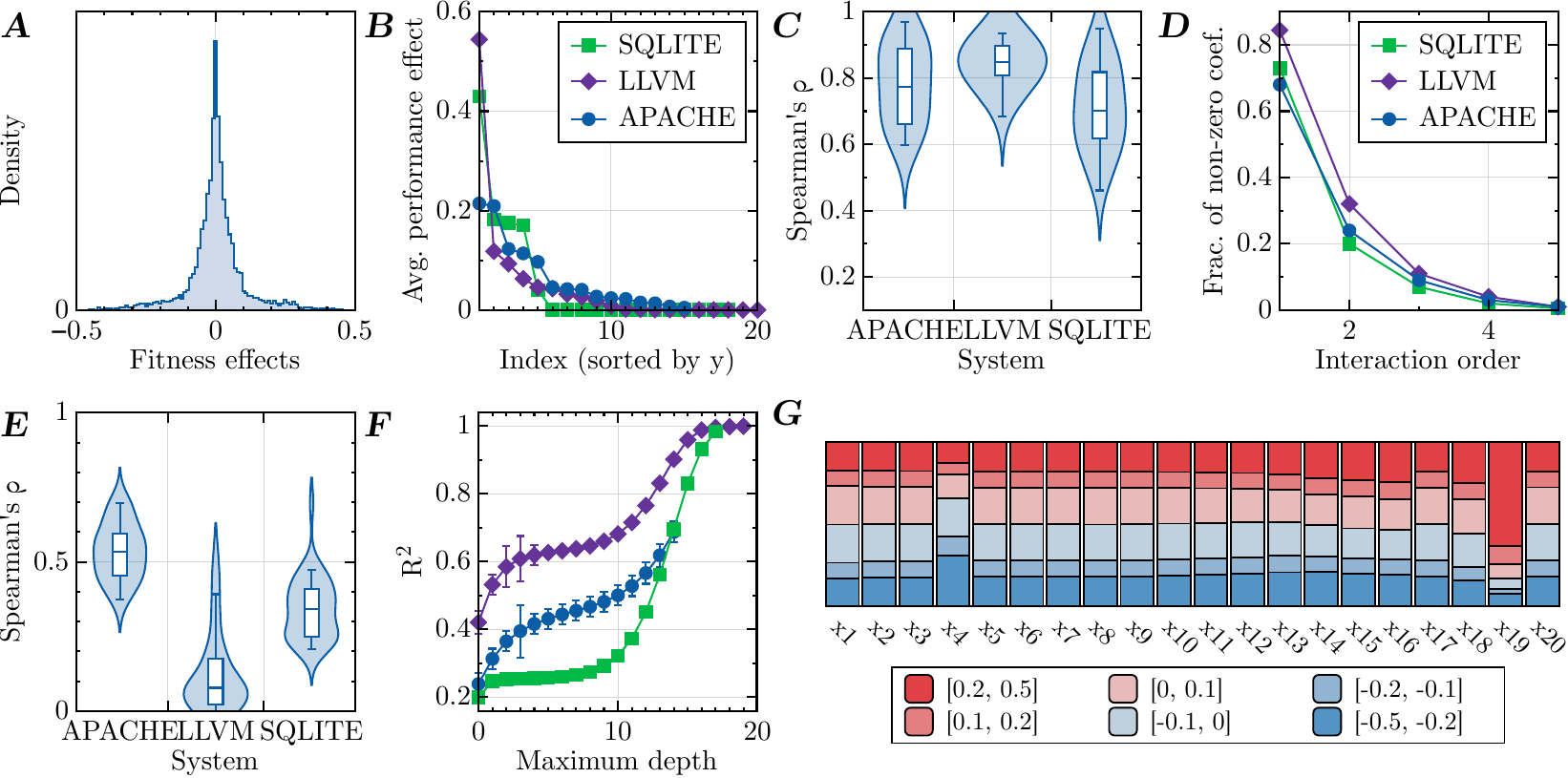}
    \caption{\textbf{Individual and interactive fitness effects of options.} \textbf{(A)}. The distribution of fitness effects (normalized) of altering a single option of \texttt{LLVM} in all possible configuration backgrounds. \textbf{(B)} Aggregated fitness effects (i.e., feature importance) of each option across all backgrounds. \textbf{(C)} The Spearman correlation between option's overall fitness effect (a.k.a. feature importance) across all workloads of each system. \textbf{(D)} Fraction of non-zero interaction coefficients as determined by LASSO regression. \textbf{(E)} The Spearman correlation between option's pairwise interaction effects (a.k.a. feature interactions) across all workloads of each system. \textbf{(F)} $R^2$ score of a random forest regressor in fitting the landscape when using different degrees of interactions (\texttt{max\_depth}). \textbf{(G)} The same as \textbf{(A)}, but plots all $20$ options in \texttt{LLVM}.}
    \label{fig:F4}
\end{figure}

\subsection{Fitness effects of individual options} 
\label{sec:f5}

So far, we have focused on explicit features of the configuration landscapes, whereas the inherent mechanism that shapes the observed topography is unexplored. We start interrogating this by examining the fitness effect of altering every single option (i.e., a \textit{mutation}) in a system. While on the surface this is similar to the traditional \textit{performance-influence} analysis~\cite{SiegmundGAK15}, prior works can only report a macro effect over all configurations and rely on predictive models that could introduce undesired modeling bias~\cite{OtwinowskiP14}. Here with combinatorially complete landscapes and \our, we were able to directly measure every single mutation's fitness effect with every possible combination of other options (a.k.a. configuration background), thus offering a micro-level assessment. 

For example, in panel A of~\pref{fig:F4}, we plotted the distribution of the fitness effects of altering a single option in \texttt{LLVM} across $2^{19} = 524,288$ configuration backgrounds. It is intriguing to see that the fitness effects of a single mutation can be both detrimental ($50.41\%$) and beneficial ($49.59\%$) in a substantial number of configuration backgrounds. This phenomenon also exists for other options in \texttt{LLVM} (see panel E of~\pref{fig:F4}) and other systems. Yet, when averaging the effects across all backgrounds, we found that not all options have prominent effects on fitness (panel B of~\pref{fig:F4}). Notably, for \texttt{SQLite}, only $5$ out of $16$ ($31.25\%$) options have statistically significant fitness effects. For the other two systems, only $4$ (\textsc{LLVM}) or $5$ (\textsc{Apache}) options have mean effects larger than $0.05$ (normalized value). Considering that we had already employed pre-selection on the options (\pref{sec:setup}), this result would imply that in practice the majority of options may not have a fitness impact. Such finding is in line with the reports from~\cite{XuJFZPT15}, where the authors found that in practice up to $54.1\%$ of configuration options are rarely touched by users, and just $1.8\%\sim7.8\%$ are frequently used by over $90\%$ of users.

While the above analyses were based on individual workloads, we further investigated whether the fitness effects of each option are consistent across different workloads of a system. Panel C of~\pref{fig:F4} shows a high Spearman's $\rho$ in this for the majority pairs of workloads in each system, with median being $0.8468$, $0.7121$, and $0.7826$ for \textsc{LLVM}, \textsc{SQLite}, and \textsc{Apache}, respectively. This indicates a good level of consistency in the option's fitness effects.

\begin{tcolorbox}[breakable, title after break=, height fixed for = none, colback = mybackground, boxrule = 0pt, sharpish corners, top = 4pt, bottom = 4pt, left = 4pt, right = 4pt, toptitle=2pt, bottomtitle=2pt]
    \textbf{F5:} Only a small proportion of options have a significant impact on system performance, and the fitness effects of altering a single option are highly context-dependent---The same mutation can result in totally different performance outcomes (e.g., sign-changing) depending on other options; The influential options are generally consistent across workloads.\\
    \textbf{Implications:} While prior performance-influence analysis can provide valuable insights, they should be interpreted as a macro view to answer questions like ``\textit{In general, which option is the most important?}''. Yet, in cases we encounter per-instance decision, e.g., ``\textit{should I turn this option on for this current configuration?}'', more granular per-instance analysis should be performed with FLA. In addition, the small proportion of influential options offers a potential for more efficient configuration tuning, e.g., by pruning the search space to concentrate the search resources on the most important options. Such a strategy has been previously shown to be effective in the related field of hyperparameter optimization for ML models~\cite{WangFT20,WistubaSS15}. Moreover, while in general, the measured feature importance is consistent across workloads, we do warn that there are also outliers in the distributions, and the trend is not perfectly preserved. This poses potential threats to transfer-based methods that assume constant feature importance like \cite{JamshidiVKS18}. 
\end{tcolorbox}

\subsection{Interactive fitness effects.} 
\label{sec:f6}

\begin{figure}[t!]
    \centering
    \includegraphics[width=\linewidth]{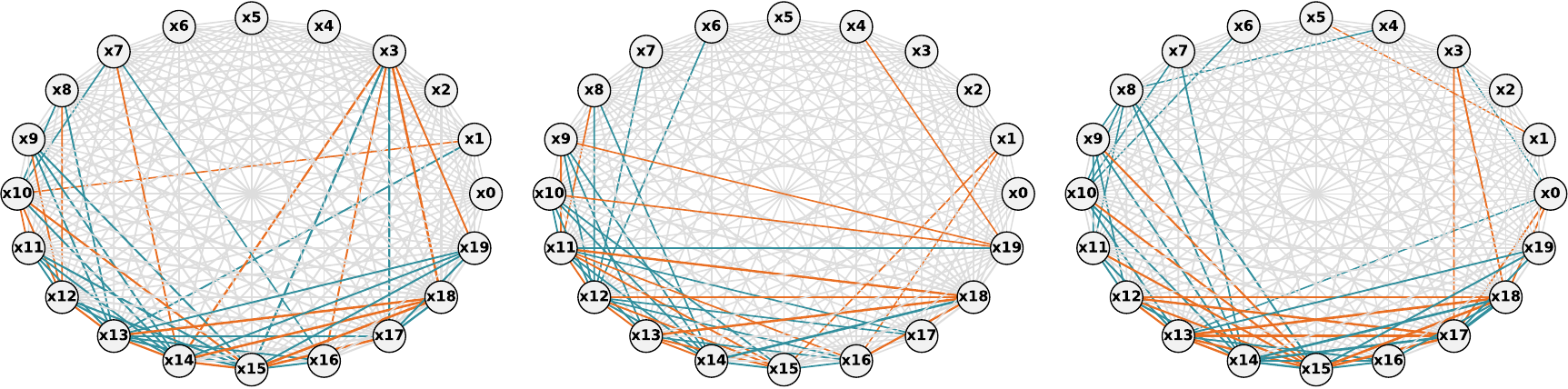}
    \caption{\textbf{Option interaction network} for three workloads of \texttt{LLVM}. Each node represents a configurable option of \texttt{LLVM} (a look-up table is available in the \href{https://zenodo.org/records/14021213}{Appendix}). \textcolor{positive}{Positive} interactions and \textcolor{negative}{negative} interactions are colored differently. Non-statistical significant interactions are colored as light grey.}
    \label{fig:interaction}
\end{figure}

In addition to individual fitness effects of options, interactions can also exist among them. From a landscape perspective, we define interaction as the difference between the collaborative fitness effect of two options and the additive sum of their individual effects. According to the sign of this difference, interaction can be further classified as negative or positive. Using \our\, we were able to precisely calculate all pairwise interactions between options. The results in~\pref{fig:interaction} indicate that certain options frequently interact with others, whereas the pattern can undergo drastic change across different workloads. In fact, as shown in panel E of~\pref{fig:F4}, for all three systems, the Spearman's $\rho$ of feature interactions across workloads are considerably lower than the results for feature importance (panel C of~\pref{fig:F4}). In particular, even for \textsc{Apache}, which exhibited the highest $\rho$ among the three, the mean is merely $0.5344$.

While previous literature predominantly focused on such pairwise interactions, considering the observed rugged topography of our studied landscapes, we postulate that higher-order interactions might be more prevalent than expected. To see this, we applied LASSO regression to assess how much fitness variation is explained by interactions of each degree between configurable options, as is common in literature~\cite{JamshidiSVKPA17,PapkouGEW23,MuhlbauerSKDAS23}. We created polynomial terms for all options up to degree 5, which is the highest we could attain. After fitting the model, we determined the proportion of non-zero coefficients at each degree. As shown in panel C of~\pref{fig:F4}, while lower-order interactions play a dominant role, higher-order interactions are also non-negligible. To scale up our analysis, we alternatively used regression trees (RTs) to fit the landscapes, where the depth of the trees can naturally serve as a cheaper proxy for the degree of interactions. For artificial $NK$-landscapes, RTs can achieve $R^2=1.0$ in fitting exactly when the tree depth reaches the specified degree of interaction $k$, which indicates that they could serve as an effective surrogate for assessing the amount of high-order interactions in landscapes. For our configuration landscapes, the results in panel D of~\pref{fig:F4} show that the $R^2$ score of the RT model consistently increases with the depth of the trees even after $10$ levels, suggesting the prevalence of high-order interactions among these landscapes, and is in line with the observed abundance of local optima in \textbf{F3}.

\begin{tcolorbox}[breakable, title after break=, height fixed for = none, colback = mybackground, boxrule = 0pt, sharpish corners, top = 4pt, bottom = 4pt, left = 4pt, right = 4pt, toptitle=2pt, bottomtitle=2pt]
    \textbf{F6:} Certain options exhibit frequent interactions with others, yet the pattern of these interactions varies substantially across different workloads. Higher-order interactions are prevalent in studied landscapes, contributing to a significant portion of the variance in performance.\\
    \textbf{Implications:} While such high-order interactions have been overlooked in many prior works, here our results suggest that they actually play an important role in shaping the rugged landscape topography as in \textbf{F3} and performance modeling. The drastic change in interaction patterns across workloads also violates the assumptions of existing transfer learning methods. Despite this, considering the distributional similarity in local optima in different workloads (\textbf{F4}), there could be deeper mechanisms that underlie the configuration landscapes. In such cases, learning latent deep features across workloads~\cite{LongC0J15}, in addition to fitness distributions, feature importance/interactions employed by current methods, may lead to more robust models. 
\end{tcolorbox}

\subsection{A recap with algorithmic experiments} 
\label{sec:alg}

We have now obtained a holistic picture of the topography of the studied configuration landscapes--from fitness distributions, to prominent regions, local optima/global optimum and feature importance/interactions. We end our analysis with actual algorithmic experiments on performance modeling and optimization and discuss how the results can be related to our findings above.  

\subsubsection{Performance modeling.} We first evaluate the capability of typical performance models in accurately predicting the fitness of configurations in our studied landscapes. To this end, we considered two prominent ML models: The first one is a random forest (\texttt{RF}) regressor~\cite{Ho95}, a strong tree-based ensemble model, which has been widely used in performance modeling in both software configuration~\cite{GuoCASW13,MartinAPLJK22,GongC22} and other neighboring fields~\cite{HutterXHL14,SiemsZZLKH20,KleinDHLG19,GrinsztajnOV22,McElfreshKVCRGW23}. We then considered a DNN, specifically the DeepPerf model proposed in~\cite{HaZ19}. It follows a simple feedforward nets structure, which is the most popular DNN architecture in performance modeling~\cite{GongC24}. While we are aware of other timely, advanced DNNs for this purpose, we leave a comprehensive benchmarking on them for future works, as it is not the primary goal of this work. For each landscape, we randomly sampled $1\%$ configurations as the training set. To allow for fair results, we tuned the hyperparameters of both models using the Tree-structured Parzen estimator (TPE)~\cite{BergstraBBK11} implemented in Optuna with an evaluation budget of $100$. We then trained the models with the optimal hyperparameters on the training set, and applied them to infer the fitness of the remaining configurations. 

\begin{figure}[t!]
    \centering
    \includegraphics[width=\linewidth]{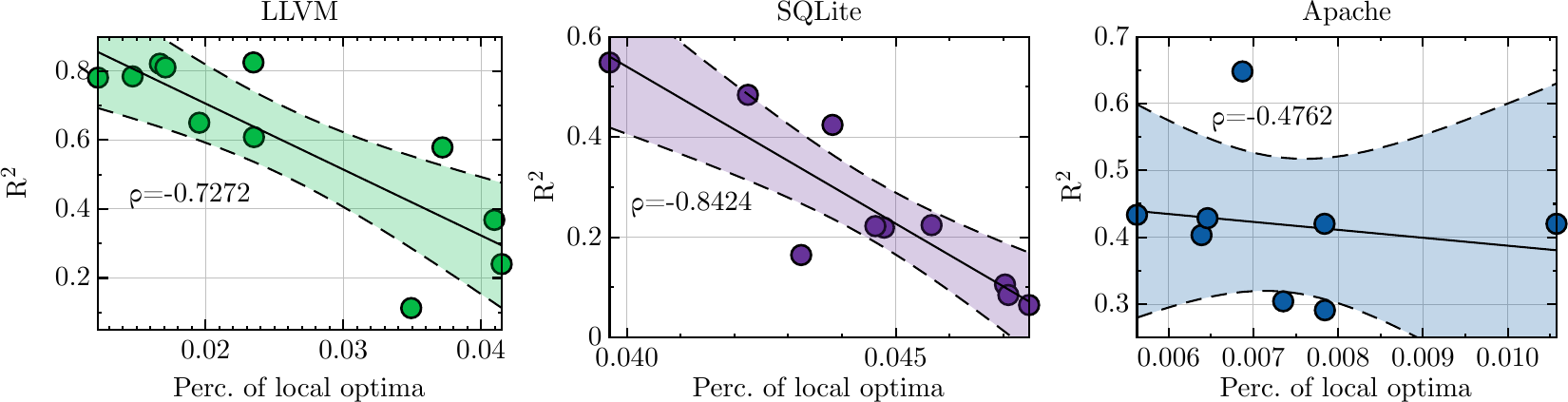}
    \caption{\textbf{The predictive power of performance models in fitting a landscape is highly dependent on its ruggedness.} Here each point represents a workload, and the prevalence of local optima in the landscape is plotted against the $R^2$ score of a random forest regressor in fitting the fitness data. A linear regression fit line, the $95\%$ confidence interval, as well as Spearman's $\rho$, are also shown in the plots.}
    \label{fig:predictability}
\end{figure}

\begin{figure}[t!]
    \centering
    \includegraphics[width=\linewidth]{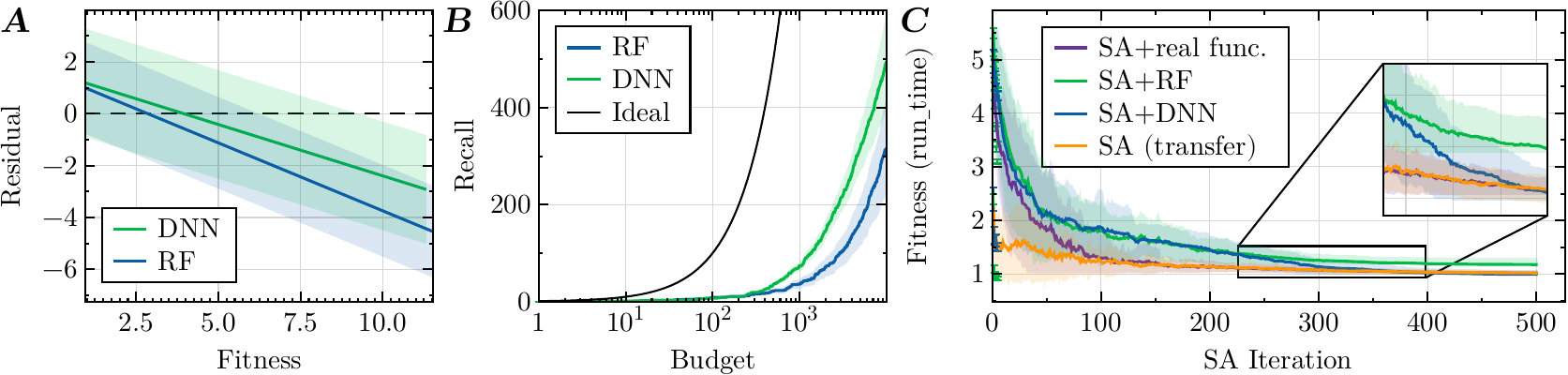}
    \caption{\textbf{(A)}. Residual plots of the fitness predictions of \texttt{RF} and \texttt{DNN} models on a random workload of \texttt{LLVM}. Linear regression lines with $95\%$ confidence interval (shading) are displayed. \textbf{(B)}. Recall of the top-$1,000$ configurations in the landscape selected by the two models under different budgets. Optimal represents a theoretical model that always predicts the true top $N$ configurations. Shading represents $95\%$ confidence intervals across all workloads of \texttt{LLVM} and $100$ runs. \textbf{(C)}. The convergence trajectory of simulated annealing (SA) on the \texttt{LLVM} landscape with different fitness functions (``real func.'', \texttt{RF}, or \texttt{DNN}, with same initialization) or with initializations with the optimal configuration from similar workload (``transfer''). Shading represents $95\%$ confidence intervals across all workloads and $100$ runs.}
    \label{fig:alg}
\end{figure}

\pref{fig:predictability} depicts the $R^2$ achieved by \texttt{RF} on test data of each landscape as a function of the number of local optima. Intriguingly, we observed a clear trend where the performance of \texttt{RF} is highly dependent on the abundance of local optima (thus ruggedness) of the landscape. This then directly relates \textbf{F3} with performance modeling since the results here suggest that local optima could also be one of the main challenges for accurately modeling the landscape, in addition to configuration tuning. In $25$ out of $32$ scenarios, the \texttt{DNN} could achieve better performance compared to \texttt{RF}, with higher $R^2$ scores by a margin of $0.023\sim0.052$. The strong dependence of modeling performance on landscape ruggedness is also observable for \texttt{DNN}. On average, the two models yielded moderate $R^2$ scores of $0.529$ and $0.547$, respectively.

However, across all scenarios, even for the most benign landscapes, we observed strong prediction biases: As depicted in panel A of~\pref{fig:alg}, both models tend to overestimate the fitness of configurations with low fitness whereas underestimate those with high fitness, though this is less prominent for \texttt{DNN}. For other landscapes, such biases also exist, with the slope of the residual regression line ranging from $-0.648\sim-0.214$. Considering that fitnesses that are very high or very low are likely to contain contributions from higher-order interactions with positive or negative effects, respectively~\cite{OtwinowskiP14}, the power of these models in capturing the strong interactions as seen in \textbf{F6} may not be as strong as we expected.

Still, we can hope that these models can at least properly learn the ranks of the configurations, such that they can be useful in selecting the prominent configurations in the landscapes~\cite{NairMSA17}. To validate this hypothesis, we tasked the models with identifying the most fit configurations for a given budget $N$, so that each model would rank all configurations and select the top-$N$ candidates. We then calculated the recall as the proportion of the true top-$1,000$ that is represented in the model's predicted top-$N$ configurations. The results in panel B of~\pref{fig:alg} show that for every budget tested, the \texttt{DNN} model had better recall compared to \texttt{RF}. However, both models yielded a recall of only $0.04\sim0.07$ for $N=1,000$, suggesting that they could hardly rank the top configurations in the landscape. If the budget increases to $N=10,000$, a recall of around $0.3\sim0.5$ could be achieved for predicting the top-$1,000$. When also considering the moderate $R^2$ scores yielded on the landscapes, these models were likely to capture the general topography of the landscapes, but the details in local regions were distorted. 

\subsubsection{Performance optimization.} Instead of directly selecting the top configurations, another useful application of performance models is to serve as cheap surrogates for guiding configuration tuning. However, in light of the above results, using such biased models could also be risky, as they may drive convergence toward misleading configurations that are incorrectly evaluated as optimal. We explored this with simulated annealing (SA), an established metaheuristics for black-box optimization, which has been previously employed for configuration optimization~\cite{GuoYQ10,DingLQ15}

Specifically, we performed SA with three different fitness functions: $i)$ the real measured performance values, and the predicted performance by $ii)$ \texttt{RF} and $iii)$ \texttt{DNN}. For all runs, we set the initial temperature to $1,000$, with a cooling rate $\alpha=0.99$. We first performed $10,000$ independent SA runs for each case, and we found that the variance in the final performance of SA was typically lower than $10\%$ of the corresponding mean. This implies that the algorithm is robust to random initializations, echoing the wide spread of the prominent configurations and local optima in the landscapes as discussed in \textbf{F2} and \textbf{F4}. We thereby grounded our further analysis on the same initialization configuration for all cases with $1,000$ independent runs. Moreover, to explore whether transferring prominent configurations from similar workloads could expedite optimization, we included an additional case by initializing SA from a random top-$1\%$ configuration of a different workload that has the highest Spearman's $\rho$ in fitness distribution with the target workload.

Panel C of~\pref{fig:alg} depicts the optimization trajectory of different SA implementations. Compared to the base case, warm-starting significantly accelerated the convergence. Yet, we noted that while the transferred configurations are among the top-$1\%$ in their original workloads, there is considerable performance degradation when applying to the target workloads, despite the high overall fitness correlation between the two landscapes. This observation is consistent with our previous findings regarding the distribution of prominent regions in \textbf{F2}. In addition to this, we found that warm-starting with transferred configurations did not lead to performance improvement over the base SA (Wilcoxon signed-rank test $p>0.05$), which conforms with the large shifts in global optima locations between workloads as observed in \textbf{F4}.

As for the surrogate-guided SA, during the initial $250$ iterations, the \texttt{RF}-guided and \texttt{DNN}-guided SA exhibited similar performance, and both models converged slower compared to the base SA. The reason for this could be the distortion in the predicted landscape structure, which could lead to misleading information for the optimizer. Yet, after $250$ iterations, the \texttt{DNN}-guided SA converged to a superior performance level compared to the \texttt{RF} counterpart, which is even comparable to the base SA (Wilcoxon signed-rank test $p>0.05$). Considering that the \texttt{DNN}-guided SA does not require expensive fitness evaluations during the optimization process, this result is highly inspiring.

\section{Threats to Validity}
\label{sec:threats}

\paragraph{Threats to interal validity}

First, our case study relies on benchmark measurements of configuration performance that are susceptible to measurement bias. We mitigated this threat by repeating each measurement for $10$ times, and the width of the $95\%$ obtained confidence interval is typically $<10\%$ of the measured mean. Thus, we do not expect such bias to significantly affect our results. Second, we made engineering decisions in the selection of configuration options, which can potentially bias the results. Yet, we argue that the procedures we adopted in option selection aim to imitate the real-world engineering practices, i.e., concentrating on the options that are most likely to influence the system performance. This would allow our findings to be more relevant for building successful performance optimizers/learners. Finally, the exact landscape characteristics may vary depending on the specific FLA methods implemented within \our. For example, there are other ways to assess landscape ruggedness~\cite{VisserK}. Note that \our\ is a general framework highly flexible to accommodate any FLA methods if necessary. In fact, we indeed performed experiments with other alternatives but the results were not reported here. This is because they yielded slightly different numerical results while our main findings are consistent. We contend that the current \our\ implementation can report the most accessible and reliable results for practitioners.

\paragraph{Threats to external validity}

This can be related to how our findings can be generalized to other context. As a case study, our analysis focused on three large systems that are widely used and extensively studied in the literature to cover a broad range of application and development scenarios (see~\pref{sec:setup}). While we observed certain commonalities in landscape topography—such as the abundance of local optima—we acknowledge that different types of software can exhibit
different characteristics. Therefore, our findings may not be directly generalizable to all configurable software systems. However, we argue that GraphFLA is designed as a general-purpose black-box FLA framework and it can be easily adapted to other systems.
The FLA perspective advocated by this work can be leveraged to analyze and understand a broader set of black-box systems and to draw new insights specific to different application domains, which are otherwise impossible. In addition, for workloads selection, we have chosen a diverse set of workloads to cover different system usage scenarios (see~\pref{sec:setup}). Although there might be additional workload characteristics, our results demonstrate already for varied landscape topographies across workloads. Thus, further variations could only strengthen our message.

%!TeX root=main.tex

\section{Conclusion}
\label{sec:conclusion}

Understanding the intricate relationship between software configurations and system performance has been a long-standing goal in software engineering, yet traditional performance analysis methods are unable to interrogate the spatial topography of the configuraiton-performance mapping. In this paper we have demonstrated a prominent new way for advancing this understanding by modeling it as a high-dimensional fitness landscape. Using our developed \our\ framework and $32$ configuration landscapes constructed from different workloads of $3$ real-world software systems, we eventually arrived at $6$ main findings that could lead to new insights for both configuration tuning and performance modeling on these systems, and some of them are previously inaccessible due to the lack of spatial information. We hope this study will open a new avenue towards the understanding of configurable software systems, for which \our\ can be easily adapted to other systems and extended to incorporate more types of analysis. Also, our collected performance data can serve as a valuable testbed for further research, enabling benchmarking and modeling at an unprecedented scale.

\vspace{.5em}
\noindent \textbf{Artifacts availability.} Source code for \our\ is available via \href{https://anonymous.4open.science/r/GraphFLA-68E4}{this link}, while our collected performance data can be accessed via \href{https://drive.google.com/drive/folders/1KcmxJLVslIFFuU-Y50MnpaP9VBRfjI8H?usp=sharing
}{this link}.

\clearpage

\bibliographystyle{ACM-Reference-Format}
\bibliography{lon}
% \clearpage
% \appendix
% \input{sections/appendix}

\end{document}